\documentclass[preprint]{aastex}
\usepackage{emulateapj5,apjfonts}  

\submitted{To appear in ApJ Letters, October 2001}

\begin{document}

\title{Evidence for Jet Domination of the Nuclear Radio Emission
in Low-luminosity Active Galactic Nuclei}

\author{Neil M. Nagar\altaffilmark{1}, Andrew S. Wilson\altaffilmark{2},
	Heino Falcke\altaffilmark{3}}
\altaffiltext{1}{Arcetri Observatory, Largo E. Fermi 5, Florence 50125, 
     Italy; neil@arcetri.astro.it}
\altaffiltext{2}{Department of Astronomy, University of Maryland,
		 College Park, MD 20742; wilson@astro.umd.edu; 
                 Adjunct Astronomer, Space Telescope Science Institute}
\altaffiltext{3}{Max-Planck-Institut f\"{u}r Radioastronomie,
                 Auf dem H\"{u}gel 69,
                 53121 Bonn, Germany; hfalcke@mpifr-bonn.mpg.de}

\received{2001 June 28}
\accepted{2001 August 21}

\begin{abstract}

We present simultaneous, sub-arcsecond ($\leq$~50~pc) resolution
5~GHz, 8.4~GHz, and 15~GHz
VLA{\footnote{The VLA is operated by the National Radio Astronomy Observatory,
a facility of the National Science Foundation operated under cooperative
agreement by Associated Universities, Inc.}}
observations of a well-defined sample of sixteen low-luminosity active
galactic nuclei (LLAGNs).
The radio emission in most of these nuclei does not show the rising spectrum
(0.2$\,\lesssim\,s\,\lesssim\,1.3$, $L_\nu\,\propto\,\nu^{s}$) expected from 
thermal electrons in an
advection dominated accretion flow (ADAF) with or without weak to 
moderately-strong outflows.
Rather, the flat radio spectra are indicative of either
synchrotron self-absorbed emission from jets, convection-dominated
accretion flows (CDAFs) with L$\,\gtrsim\,10^{-5}\,L_{\rm Edd}$,
or ADAFs with strong ($p\,\gtrsim\,0.6$) outflows.
The jet interpretation is favored by three factors: a)~the detection of
pc-scale radio extensions, morphologically reminiscent of jets, in the five
nuclei with the highest peak radio flux-density; b)~the domination of
parsec-scale jet radio emission over unresolved `core' emission in the
three best-studied nuclei; and
c)~the lack of any clear correlation between radio spectral shape and
black hole mass as would be expected from the dependence of the radio
turnover frequency on black hole mass in ADAF and CDAF models.
A jet domination of nuclear radio emission implies significantly lower
accretion rates in ADAF-type models than earlier estimated from core 
radio luminosities.

\end{abstract}


\section{Introduction}

The detection of optical broad-emission-lines \citep{hoet97b}
and high brightness-temperature ($\gtrsim~10^8$~K) radio cores
\citep{bieet00,junbir95,falet00,naget01} in the nuclei of a
well-defined sample of nearby bright galaxies (Ho, Filippenko, \& Sargent 1997a)
indicates that at least 20\% of all nearby bright galaxies have
an accreting massive black hole.
These nearby nuclei have been christened low-luminosity active
galactic nuclei or LLAGNs.
Their low nuclear luminosities require either very low accretion
rates \citep[$\sim$10$^{-8}\,$L$_{\rm Edd}$; e.g.][]{falbie99}
or radiative efficiencies (the ratio of radiated energy to accreted
mass) much lower than the typical value of
$\sim$10\% \citep[e.g.~Chapter~7.8 of][]{fraet95} assumed for powerful AGNs.
This has led to renewed interest in spherical accretion models 
which produce low radiated luminosities, e.g.
advection-dominated accretion flows \citep[ADAFs;][]{naret98b},
which may have associated outflows \citep{blabeg99}, and
convection-dominated accretion flows 
\citep[CDAFs;][]{stoet99,naret00,quagru00}. 

The sub-parcsec radio emission in LLAGNs may originate in an ADAF
or CDAF inflow, and the predictions of these models are discussed
in the following section.
The sub-parsec radio emission may alternatively originate in
synchrotron radiation from discrete plasma components or from the base 
of a continuous jet ejected from the central engine.
In the former case, the presence of different self-absorption frequencies for 
individual components results in a flat spectral-shape 
($s\,\sim\,0$; $L_\nu\,\propto\,\nu^s$) for
the overall system \citep[e.g.][]{mar88}.
In the latter case, i.e. for relativistic electrons at the base of a 
continuous, freely expanding jet \citep{blakon79}, the variation of the
electron density (n$_{\rm e}\,\propto\,d^{-2}$), electron temperature
($T_{\rm ej}\,\propto\,d^{0}$), and magnetic field ($B\,\propto\,d^{-1}$), 
where $d$ is the distance along the jet axis, results in a flat overall radio 
spectrum.
Slightly inverted spectra (up to $s\,\simeq\,0.2$) may result from the
bulk acceleration of the jet plasma \citep{fal96}, and 
even higher (temporary) values of $s$ may be measured during radio outbursts
\citep[e.g.][]{hoet99}.
On larger scales the synchrotron emission from the ejecta or jet becomes
optically-thin ($s\,\simeq\,-0.7$). Thus the spectral index of the
overall synchrotron emission from a `jet' is expected to be between 0.2 and
$-$0.7, depending on the relative contributions of the base and extended
components.

\section{Radio Spectral Predictions of ADAFs and CDAFs}

In their simplest form ADAF and CDAF self-similar spherical models invoke 
standard $\alpha$ viscosity, a two-temperature thermal plasma, and a fixed 
ratio ($\beta$) of magnetic to gas pressure. 
Let us first use the analytic scaling-law approximations of 
\citet[][hereafter M97]{mah97} to 
estimate the expected radio spectral index in a self-similar flow.
If we follow the derivation in Sec.~4.1 of M97 using
magnetic field $B\,\propto\,r^{-c}$, 
electron temperature $T_e\,\propto\,r^{-a}$, 
and synchrotron emission factor $x_M\,\propto\,r^{-d}$ 
then the radio emission is self-absorbed, with spectral index
$s\,\simeq\,\frac{5a+2c+2d-2}{2a+c+d}$
at frequencies below a critical frequency $\nu_p$. 
Here $r$ is the radius in units of the Schwarzschild radius and
$c$ is 5/4 in an ADAF \citep{naryi95} and 3/4 
in a CDAF \citep{naret00}.
When $L\,\simeq\,10^{-4}L_{\rm Edd}$, the dominance of synchrotron cooling 
forces $a$ to 0 over the radio emitting region ($\lesssim$10$^3\,$r) in an
ADAF or CDAF \citep[see][]{naryi95}. 
However, when $L\,\lesssim\,10^{-7}L_{\rm Edd}$ the electrons are
adiabatically compressed so that 
$a\,\simeq\,0.5$ in an ADAF \citep{naret98a} and
$a\,\simeq\,1$ (i.e. virial) in a  CDAF \citep{naret00}.
The value of $d$ is more difficult to estimate. M97 find
$d\simeq$1/15 in ADAFs with $L\,\simeq\,10^{-4}L_{\rm Edd}$
(their Appendix B). Solving equation B1 of M97 with typical values of
$T_e(r=1)= 10^9-10^{10}$~K and $m{\dot{m}}= 10^{-1}-10^{4}$  shows that for 
an ADAF 
$d\,\simeq\,-0.15$ when $a\,=\,0$ and 
$d\,\simeq\,-0.6$ when $a\,=\,0.5$.
Here $m$ is the black hole mass in solar masses and $\dot{m}$ is the 
accretion rate in units of the Eddington rate (assuming 10\% radiative 
efficiency). 
If we simplistically modify the form of $\rho$ and $B$ in an ADAF 
(equations 5 of M97) only in the power law dependence of $r$ 
to convert them to the equivalent expresssions for a CDAF
(see Narayan et al. 2000 for some justification of this),
then we can solve the equivalent of M97's equation B1 to 
find that for a CDAF 
$d\,\simeq\,-0.2$ when $a\,=\,0$ and 
$d\,\simeq\,-1.6$ when $a\,=\,1$.
Thus, for an ADAF one expects 
0.2$\,\lesssim\,s\,\lesssim\,1.1$
and for a CDAF one may expect
$s\,=\,-1.8$ when $L\,\simeq\,10^{-4}L_{\rm Edd}$ and 
$s\,=\,1.1$  when $L\,\lesssim\,10^{-7}L_{\rm Edd}$.
Given the many approximations in this analytical method the above results 
should be considered only roughly indicative.
ADAFs with outflows{\footnote{CDAFs are considered 
separate from ``outflows'' as their convective eddies are not unbound
\citep{naret00}.}}
have density profile 
$\rho\,\propto\,r^{-\frac{3}{2}+p}$ \citep{blabeg99}
with $p$ varying from 0 (no outflow) to 1 (strong outflow).
The radio emission from the accretion flow corresponds to the ADAF case for
$p\,=\,0$ and to the CDAF case for $p\,=\,1$; radio emission from the outflow
has not been modeled and is not considered in this paper.

More accurate numerical modelling of the radio emission from an
ADAF gives $s\,=\,0.4$ when
$L\,\simeq\,10^{-4}L_{\rm Edd}$ (M97) and 
$s\,\simeq\,1$ when $L\,\lesssim\,10^{-7}L_{\rm Edd}$    
with $s$ decreasing but still positive as $p$ increases
from 0 to 0.6 \citep{naret98a,quanar99}. 
To our knowledge, there has been no published work dealing explicitly with
numerical modelling of the radio emission from a CDAF.
In summary, the radio emission below $\nu_p$ from the accretion inflow of
an ADAF or CDAF is expected to have a moderately  to highly inverted 
spectrum except for 
a)~CDAFs in systems with $L\,\gtrsim\,10^{-5}\,L_{\rm Edd}$ and 
b)~ADAFs with strong ($p\,\gtrsim\,0.6$) outflows.

The inflow is optically thin at frequencies greater than 
\begin{equation}
\nu_p\,=\,1.3\,\times\,10^{-4}\,\alpha^{-1/2}(1-\beta)^{1/2}x_Mm^{-1/2}\dot{m}^{1/2}T_e^2\,r_{\rm min}^{-5/4}
\label{eqnnu_p}
\end{equation}
so the radio emission falls off exponentially above $\nu_p$ (M97).
For ADAFs without outflows $T_e$ and $x_M$ are relatively independent
of $m$ for $m\,=\,10^7$--$10^9\,$M$_{\sun}$ (Fig.~2 of M97).
Thus, for typical values of $\alpha$ (0.3), $\beta$ (0.5), and $\dot{m}$
($10^{-4}$--$10^{-6}$), one has $\nu_p\,\simeq\,10^{15-16}m^{-1/2}$, i.e. the 
turnover frequency is greater than 15~GHz for all black holes considered here. 
In ADAFs with outflows and in CDAFs the flatter density profile results 
in a decrease in the values of $T_e$ and $B$ near $r\,\simeq\,1$, and 
consequently a decrease in $\nu_p$ (and L$_{\nu_p}$); the introduction of
a moderately strong ($p\,=\,0.6$) outflow to an ADAF can lower the value of 
$\nu_p$ by almost two orders of magnitude \citep{quanar99}.

\section{Observations and Data Reduction}

Sixteen of the 96 nearest (D~$\leq$ 19~Mpc) LLAGNs from the Palomar sample 
\citep{hoet97a} have a highly compact ($\leq$~2~mas), 
high brightness-temperature ($\gtrsim\,10^8$~K) radio core
\citep[see][]{falet00,naget01}. 
These sixteen LLAGNs, listed in Table~1,
were observed with the Very Large Array (VLA) at 5~GHz (6~cm),
8.4~GHz (3.6~cm), and 15~GHz (2~cm).
The observations were made on September 5 and 
September 10, 1999, while the VLA was in
``A''-configuration (see Thompson et al.\ 1980).
Each LLAGN observation was sandwiched between two observations
of a nearby phase-calibrator with typical cycle times, in minutes,
of 1-7-1, 1-6-1, and 2-7-2,  at 5~GHz, 8.4~GHz, and 15~GHz, respectively.
The observations at the three frequencies are simultaneous to $\leq$30 min
for each LLAGN.

Data were calibrated and mapped using the AIPS software, following the 
standard procedures outlined in the AIPS
cookbook.
Observations of 3C~147 and 3C~286 were used to set the 5~GHz flux-density
scale, and observations of 3C~286 were used to set the 8.4~GHz and
15~GHz flux-density scales.
The 15~GHz observation of 3C~286 was made at a single (1.4) airmass, and
all 15~GHz observations were made at airmasses of 1.06 to 1.4, so
the 15~GHz flux calibration error from elevation effects is expected to be
less than 0.2\% \citep{per00}.
For this reason we did not make elevation-dependent gain corrections to
the 15~GHz data.
The VLA documentation suggests that the flux calibration at 5~GHz and 8.4~GHz
should be accurate to 1\%--2\%, and that at 15~GHz should be accurate
to 3\%--5\%; we conservatively use the higher numbers as the respective 
$2\,\sigma$ errors.
For sources with flux greater than 3~mJy, we were able to iteratively 
self-calibrate (both phase-only and amplitude-and-phase) and image the data 
so as to increase the signal-to-noise ratio in the final map.
The root mean square noise in the final uniformly weighted maps was
typically 100~$\mu$Jy, 60~$\mu$Jy, and 170~$\mu$Jy at
5~GHz, 8.4~GHz, and 15~GHz, respectively.
The resolution at these three wavelengths was typically 0{\farcs}5,
0{\farcs}27 and 0{\farcs}15, respectively.
We also made 15~GHz and 8.4~GHz maps with the same resolution (0{\farcs}5)
as the 5~GHz maps, by appropriately tapering the $(u,v)$ data.

\section{Results}
All sources except NGC~4168 at 15~GHz 
were clearly detected in initial (non-self-calibrated) maps.
The 15~GHz observation of NGC~4168 was made during very bad weather, and 
we were able to make a noisy map only after self-calibration
with a point-source model.
The newly measured flux-densities are listed in Table~1. 
The 15~GHz data are noisy because of bad weather and high humidity. 
Therefore the three nuclei for which we could not self-calibrate the 
15~GHz data 
have true 15~GHz fluxes 
somewhere between the measured values and 3~mJy.
For all but three of the objects, the radio emission at all three frequencies
is compact; a Gaussian fit to the source does not give a deconvolved size
more than half a beam-size. The three sources with detected extended structure
are all previously known to have such structure:
NGC~4278 \citep{wilet98}, NGC~4472 \citep{ekekot78}, and 
NGC~4486 \citep[M~87; e.g.][]{junbir95}.
The unresolved emission dominates the extended emission in our maps
of these three sources except in NGC~4472, which has a very weak core.
Most nuclei have roughly similar fluxes in the full resolution 15~GHz maps and
the 0{\farcs}5 resolution tapered maps (Table~1); the same is true at 8.4~GHz.
The peak flux-density in the 0{\farcs}5 resolution, 5~GHz VLA maps is
$\sim$0.8--2.1 times the total (but not necessarily core) flux in the central
$\leq$20~mas of the 5~GHz (non-simultaneous) VLBA maps for all sources
which were observed in our June 1997 and April 1999 VLBA runs
(column 14 of Table~1).

The variation of the core spectral index 
(from the peak fluxes in matched resolution maps) with
black hole mass is shown in Fig.~\ref{figmdo_alpha}a.
We have distinguished between black hole masses
derived directly from 
stellar-, gas-, and maser-dynamics
\citep{gebet00,ricet98} from those
inferred from central velocity dispersions
\citep[using the relationship derived by][]{gebet00}
and galaxy bulge masses
\citep[using the relationship derived by][]{ricet98}.
Only NGC~3031 (M~81) and NGC~4772 consistently show the highly inverted radio
spectrum expected in ADAF models with or without weak to moderately-strong
outflows. The core radio emission from the latter galaxy is probably dominated
by extended emission (see above), and as discussed below the radio emission 
from the former is probably from a jet.
Apart from NGC~4472, the elliptical 
galaxies have a similar spectral shape above and below 8.4~GHz. 
Most of the non-ellipticals have a spectrum which falls more rapidly 
at frequencies above 8.4~GHz than below (Fig.~\ref{figmdo_alpha}b), with
NGC 3718 and NGC 4258 being the exceptions.
When high ($\leq$2{\arcsec}) resolution 2-10~keV X-ray luminosities
are available \citep[e.g.][]{hoet01}, the ratio of
$\nu$L$_\nu$ between X-ray and radio is $\sim\,10^{1-3}$, suggestive of
strong outflows in an ADAF scenario (Di Matteo, Carilli, \& Fabian 2001). Interestingly, this 
ratio is $\gtrsim$100 for the Seyfert~1s and $\lesssim$100 for the other nuclei.

\section{Discussion}

Very low accretion rates (perhaps due to convection 
or strong outflows) may cause $\nu_p$ to fall close to 5--15~GHz
for the objects in the sample.  Such a scenario is supported by the 
evidence for turnover frequencies in the 10--30~GHz range for a few 
ellipticals \citep{dimet01}.
If this is the case, then eqn.~\ref{eqnnu_p} implies lower values of $\nu_p$ 
for more massive black holes. That is, within our sample we would 
expect nuclei with less massive black holes to have more inverted spectra 
than nuclei with more massive black holes.
However, even though we sample more than two orders of
magnitude in $m$, Fig.~1 does not support such a trend.
Therefore, unless non-ellipticals have different micro-physical parameters,
or higher accretion rates, or a different accretion mechanism, as compared
to ellipticals, it is unlikely that a turnover frequency in the  5--15~GHz 
range is the cause of the observed flat radio spectrum in most of the sample.
If $\nu_p\,>\,15$~GHz for most nuclei in the sample, then any inverted 
spectrum radio 
component must be dominated by other sources at 
5~GHz and 8.4~GHz (and perhaps even at 15~GHz).
One potential source is non-thermal electrons within the ADAF \citep{ozeet00}.
Significant emission from star-formation related processes can be ruled out
as the radio core has a high brightness-temperature at 5~GHz, and at this
wavelength most of the flux within the central 0{\farcs}5 is also detected 
on mas-scales (Table~1). 
On the other hand, the observed distribution of spectral indices is consistent
with the 5--15~GHz radio emission originating in synchrotron-emitting jets. 

Whether or not an accretion flow contributes to the nuclear radio emission,
the detections of what appear to be collimated pc-scale jets in the five
LLAGNs of Table~1 with the highest core flux -
NGC~3031 \citep[M~81;][]{bieet00},
NGC~4278 \citep{jonet84,falet00},
NGC~4486 \citep[M~87;][]{junbir95},
NGC~4374 \citep[M~84;][]{wroet96,naget01},
and NGC~4552 \citep[M~89;][]{naget01} -
does indicate that synchrotron emission from jets is a significant
contributor to the sub-arcsecond radio emission.
In fact, in all three sample nuclei which have been comprehensively
studied at high resolution in the radio, the radio flux from the jet
dominates that from the unresolved ``core.''
In NGC~4486 the jet component within 30~mas (2.5~pc) of the nucleus 
contributes three times the radio flux of the unresolved (1~mas x 0.2~mas) 
``core''  \citep{junbir95}.  Given that this ``core''
continues to be further resolved at higher resolutions 
(Junor, Biretta, \& Livio 1999),
and that the jet is a strong radio emitter on scales of 30~mas to 1{\arcsec},
the jet is certainly the dominant sub-arcsecond radio emitter.
In NGC~3031, which has a spectral shape consistent with an ADAF
model, sub-mas multi-epoch observations reveal that the
sub-parsec jet contributes at least three times the radio flux of the
unresolved ``core'' \citep{bieet00}.
Deep radio observations of NGC~4258 not only reveal a sub-parsec jet, but
also indicate an absence of continuum emission from the putative location
(as traced by the water-vapour maser disk) of the nucleus \citep{heret97}.

In the context of any low-luminosity spherical accretion model, the 
presence of outflows and the smaller radio luminosities
attributable to the inflow both point to accretion rates at least
an order of magnitude lower than than earlier predicted using ADAF models
\citep[e.g. 10$^{-2}$--10$^{-4}~L_{\rm Edd}$;][]{chaet00,yibou99}.
A jet can also cause considerable disruption of the high-frequency radio 
emitting region ($\sim\,1\,r\,-\,100\,r$) of the inflow.
\citet{junet99} find a wide ($\gtrsim\,60{\arcdeg}$)
initial opening angle for the (potentially relativistic) radio jet in
NGC~4486, with collimation only occurring at $\sim100\,r$.
For a moderately-strong outflow (e.g. $p\,=\,0.6$) about 25\% of the 
material accreted at 100$\,r$ is lost to the outflow by $2\,r$.
Thus it may not be accurate to model the radio emitting region
as a spherical self-similar flow in which the only effect of the outflow is 
a modification of the accretion rate and central density.

\epsscale{1.1}
\vspace{-3.3in}
\hspace{-1.0in}
\plotone{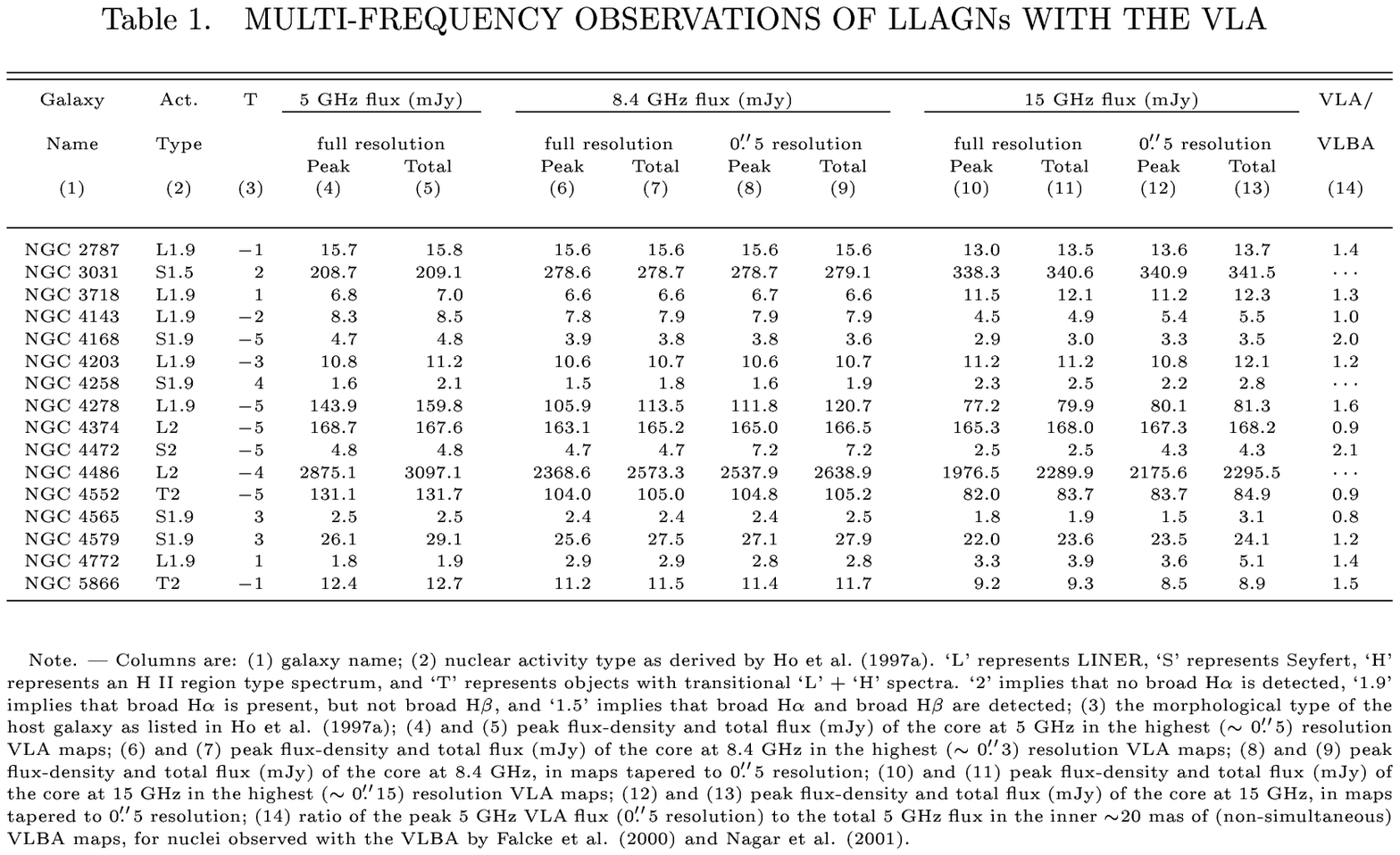}                                                                                                         

\epsscale{1.05}
\vspace{-3.7in}
\plottwo{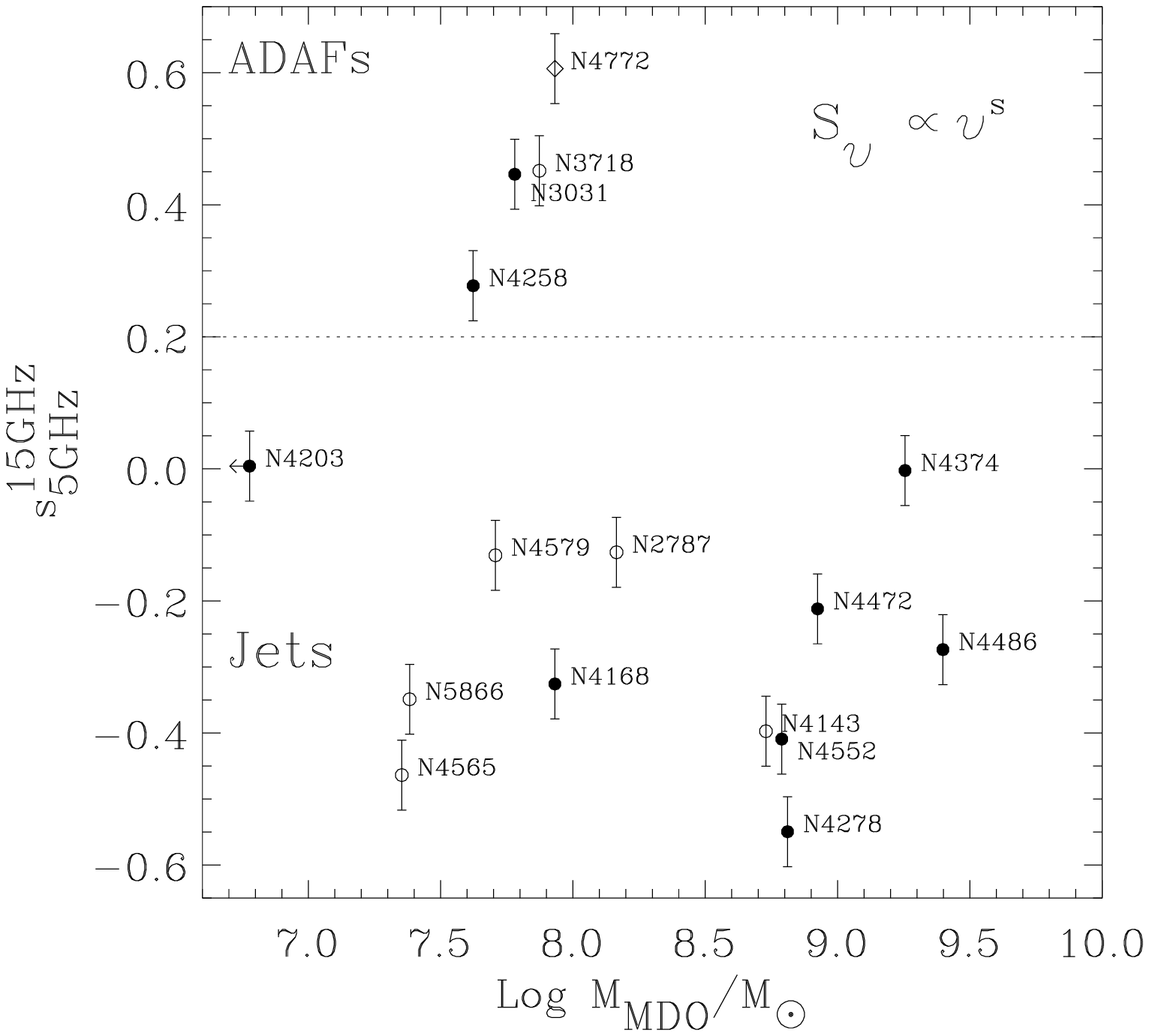}{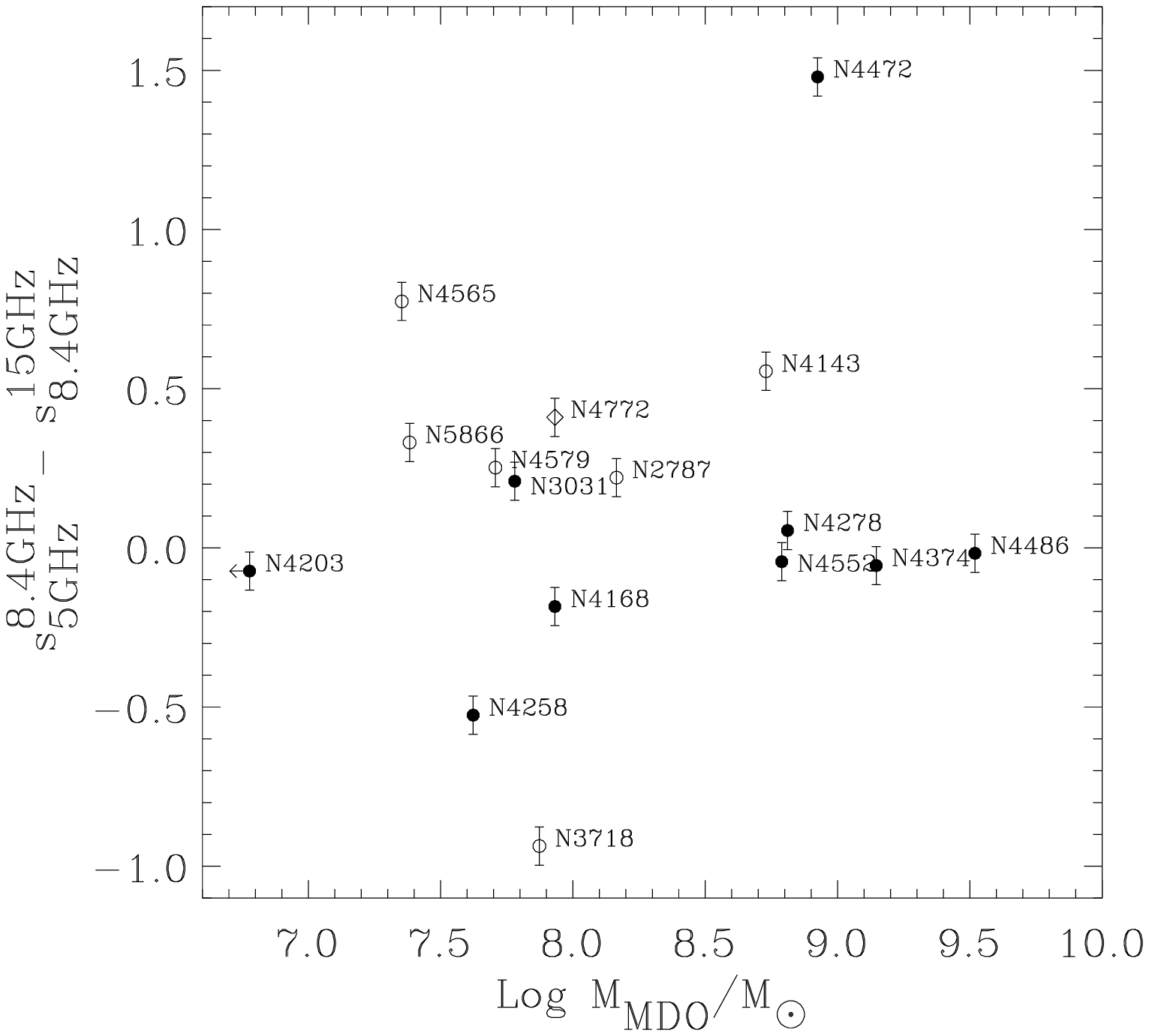}

\figcaption[fig1.ps]
{The (a)~spectral index and (b) change in the spectral index,
 between 5~GHz and 15~GHz, as a function of black hole mass.
 Nuclei with positive $y$-axis values in (b) have spectra which fall off more 
 steeply above 8.4~GHz than below. 
 Filled black circles are used for reliable black hole estimates,
 and open circles and open diamonds are used for black hole masses inferred
 from the relationships of \citet{gebet00} and \citet{ricet98}, respectively
 (see text).
 Two sigma error bars in $y$ are shown.
 \label{figmdo_alpha}
}

\end{document}